\documentclass[12pt,preprint]{aastex}
\slugcomment{UNIL-IPT-02-1}
\shorttitle{$\gamma$-ray versus UHECR emission by BL Lac objects}
\shortauthors{Gorbunov et.al.}
\begin{document}
\title{%
Evidence for a connection between the $\gamma$-ray and the highest
energy cosmic-ray emissions by BL Lacertae objects
}
\author{D.~S.~Gorbunov\altaffilmark{1,4},
P.~G.~Tinyakov\altaffilmark{1,2,5},
I.~I.~Tkachev\altaffilmark{1,3,6},
and
S.~V.~Troitsky\altaffilmark{1,7}
\altaffiltext{1}{Institute for Nuclear Research of the Russian Academy of
Sciences,
60th October Anniversary Prospect 7a, 117312, Moscow, Russia}
\altaffiltext{2}{Institute of Theoretical Physics, University of Lausanne,
CH-1015, Lausanne, Switzerland}
\altaffiltext{3}{CERN Theory Division, CH-1211 Geneva 23, Switzerland}
\altaffiltext{4}{e-mail: {\tt gorby@ms2.inr.ac.ru}}
\altaffiltext{5}{e-mail: {\tt Peter.Tinyakov@cern.ch}}
\altaffiltext{6}{e-mail: {\tt Igor.Tkachev@cern.ch}} 
\altaffiltext{7}{e-mail: {\tt st@ms2.inr.ac.ru}}
}
\date{}
\begin{abstract}
A set of potentially $\gamma$-ray--loud BL Lac objects is selected by
intersecting the EGRET and BL Lac catalogs. Of the resulting 14
objects, eight are found to correlate with arrival directions of
ultra-high-energy cosmic rays (UHECRs), with significance of the order
of $5\sigma$.  This suggests that $\gamma$-ray emission can be used as
a distinctive feature of those BL Lac objects that are capable of
producing UHECR.
\end{abstract}

The highest energy cosmic rays with energies in excess of $10^{19}$~eV
(ultra-high-energy cosmic rays [UHECRs]), observed by AGASA \citep{AG}
and Yakutsk \citep{YK} experiments, show a significant number of
clusters at angles of the order of experimental angular resolution
\citep{clusters1}. The significance of clustering is quantitatively
estimated by calculating the angular correlation function of the UHECR
events \citep{Tinyakov:2001ic}. It follows that the observed
clustering has probability of less than $10^{-5}$ to occur as a result
of a statistical fluctuation. This suggests that (1) there exist
compact sources of UHECRs and (2) the already existing data may
contain information sufficient to identify the actual sources, the
subset of cosmic rays with maximum autocorrelations being the best
choice for this purpose.

This line of reasoning was pursued by \citet{Tinyakov:2001nr} assuming
that BL Lacertae objects are relevant candidates.  Significant
correlations were found with the subset of most powerful confirmed BL
Lac objects. After assigning penalties for subset selection and bin
size adjustment, the probability of such correlation to occur by
chance in a random distribution is of order $10^{-4}$. BL Lac objects
comprise a subclass of blazars characterized by the absence of
emission lines. Blazars are thought to have relativistic jets directed
along the line of sight, while the absence of emission lines indicates
low ambient matter and radiation fields and therefore favorable
conditions for the acceleration of particles to highest energies. For
this reason, BL Lac objects may be particularly promising candidates
for UHECR sources.

It follows from both the statistical arguments \citep{Dubovsky:2000gv}
and correlation analysis \citep{Tinyakov:2001nr,Tinyakov:2001ir} that
only a small fraction of existing BL Lac objects should be capable of
producing highest energy cosmic rays. For understanding the nature of
the sources and the mechanism of UHECR emission, the key question is
which physical characteristics single out the actual emitters among
all BL Lac objects? In this Letter we propose that the strong
$\gamma$-ray emission is the feature that distinguishes UHECR sources.

There are general reasons to expect the connection between UHECR and
$\gamma$-ray emissions. Both the acceleration of particles in the
source and their subsequent propagation in the intergalactic space is
accompanied by energy losses. A substantial part of this energy is
transferred into the electromagnetic cascade and, generically, ends up
in the EGRET energy region \citep{Berezinsky,Aharonian}.  In models
involving neutrinos via the Z-burst mechanism
\citep{Fargion,Weiler,Fargion:2001pu}, and those based on very
high-energy photons \citep{Kalashev:2001qp,Neronov:2002se}, the
astrophysical accelerator must be very powerful to provide sufficient
flux of primary ultra--high-energy particles. In these models, one may
expect a strong electromagnetic radiation from the source and
substantial contribution into EGRET flux. Note that the {\em
extragalactic} cascade may get isotropized by random magnetic fields
when approaching the low energy end; this may cause smearing of point
sources and result in contributions into $\gamma$-ray background. In any
case, these arguments suggest that $\gamma$-ray emission may be an
important distinctive feature of UHECR sources\footnote{We are
grateful to A. Neronov and D. Semikoz for numerous useful discussions
of this subject.  Note that possible connection between gamma and
neutrino signals has also been discussed in \citet{Fargion2}.}.

In order to test this hypothesis we first select those BL Lac objects
that can be associated with $\gamma$-ray sources and then study their
correlations with UHECR. The most complete list of the $\gamma$-ray
sources can be found in the third EGRET catalog \citep{3EG} containing
271 object. Of these objects, 67 are identified with active galactic
nuclei (AGNs), five with pulsars, one with a solar flare, one with the
LMC, and 27 are tentatively identified with AGNs. The remaining 170
objects are unidentified.

In this Letter we do not rely on the existing EGRET identification of
objects, neither do we attempt our own object-by-object analysis.
Instead, we adopt a purely statistical approach: we take the full set
of confirmed BL Lac objects from the Veron2001 catalog \citep{Veron}
consisting of 350 objects, and we select a subsample of those that may
be associated with an EGRET $\gamma$-ray source. The selection
procedure is as follows: Point sources in the EGRET catalog are
defined as a local excess of a signal over the background. Each source
is associated with a contour containing 95\% of the signal. For each
contour, a circle of equal area is defined, with the radius $R_{95}$.
These radii are listed in the EGRET catalog. They roughly correspond
to uncertainties in the positions of the sources.  However, the 95\%
contours are often noncircular. Additional systematic errors in
position determination may be present in the case of a bright nearby
source (such cases are marked as ``confused'' in the catalog). As a
result, many well-identified sources (e.g., the Vela pulsar that is
unambiguously identified by timing) fall outside of $R_{95}$. In our
analysis, we consider an object to be associated with the EGRET source
if the angular distance between the two does not exceed $2R_{95}$. In
cases of ambiguity the nearest neighbor is taken.

According to this procedure, 14 BL Lac objects from the Veron2001
catalog are associated with EGRET sources. They are listed in Table~1.
Of these 14 objects, eight already have identifications in the EGRET
catalog, while six are newly proposed identifications. Out of eight
previously identified objects, five have the same identifications in
the SIMBAD database as is suggested by our procedure (objects marked
by asterisks in Table~1). Interestingly, in those three cases when our
procedure suggests identification different from the existing one, the
latter has a question mark in the SIMBAD database, while in five cases
when they coincide the existing identification is considered
firm. This rather good agreement with previous results gives
confidence that at least part of previously unidentified EGRET sources
listed in Table~1 should be identified with corresponding BL Lac
objects.

Since the EGRET 95\% contours are large enough to contain several
astrophysical objects, the identifications depend on the assumptions
about candidate sources.  Most previous works have concentrated on the
powerful radio quasars as possible candidates (see, e.g.,
\citet{identifications}). An approach somewhat similar to ours was
used by \citet{punsly} where correlations of EGRET catalog with X-ray
and moderate radio sources ({\it ROSAT}--Green Bank catalog) were
considered. It revealed several new identifications, large fraction of
them being BL Lac objects.

Being based on position coincidence only, the identifications proposed
in Table~1 cannot be considered as final. Instead, Table~1 should be
treated as a starting point for more detailed object-by-object study,
including EGRET intensity maps, time correlations, etc. Such an
analysis goes beyond the scope of this Letter. It is important to
note, however, that possible misidentifications in Table~1 do not
compromise our main result, strong correlation of the selected
subsample with UHECRs. Like any random factor, such misidentifications
can only diminish the correlations.

Let us now turn to correlations between the set of 14 (potentially)
$\gamma$-ray--loud BL Lac objects of Table~1 and UHECRs. In the part
concerning UHECRs, we follow the approach of \citet{Tinyakov:2001nr}
and use the set of cosmic rays with the largest autocorrelations.
This set consists of 39 AGASA events with energies $E>4.8\times
10^{19}$~eV and 26 Yakutsk events with energies $E>2.4\times
10^{19}$~eV \citep{Tinyakov:2001ic}.

The numerical algorithm used in this Letter is identical to that of
\citet{Tinyakov:2001ic,Tinyakov:2001nr,Tinyakov:2001ir}. We
characterize the significance of correlations between UHECRs and a
given set of sources at a given angular scale $\delta$ by the
probability $p(\delta)$ defined in the following way. First, we count
the number of source/cosmic-ray pairs separated by the angle $\leq
\delta$ in the real data, thus obtaining the data count
$N_d(\delta)$. We then generate a large number of random (mock) sets
of cosmic rays, taking into account actual acceptance of the
experiments in such a way that the large-scale distribution of mock
cosmic rays is uniform. On small scales we introduce autocorrelations
in mock sets since the real data are clustered. The number of clusters
added in each mock set mimics the real data, while cluster positions
are random. For each mock set, the number of source/cosmic-ray pairs
is then counted in the same way as for the real data, giving the mock
count $N_m(\delta)$. At a large total number of mock sets, the
fraction of mock sets for which $N_m(\delta)\geq N_d(\delta)$ gives
$p(\delta)$.

In the correlation analysis, we take into account possible effects of
the Galactic magnetic field (GMF) on propagation of UHECRs. We use the
spiral model of GMF with different directions of the field in the two
spiral arms and consider two cases: symmetric and antisymmetric
field with respect to the galactic plane. The details of the model and
corresponding parameters can be found in \citet{Tinyakov:2001ir}
together with further references. We assume that primary particles can
have charges of $Q=0,\pm 1$. 

In the case $Q\neq 0$, the positions of cosmic rays are corrected for
the deflections in GMF prior to counting the number of pairs with
given angular separation. For each cosmic ray there are several
possible positions after correction for GMF corresponding to different
allowed charges. For a given ray, the minimum angular distance over
the set of sources and charges determines the resulting charge
assigned to that ray. In all cases, each randomly generated set is
subject to exactly the same procedure as the real data. This
guarantees that no correlations are artificially introduced.

The results of the calculations for the charge assignments $Q=0$,
$Q=1$, $Q=0,1$ and $Q=0,\pm1$ and for two types of magnetic field
(symmetric and antisymmetric) are presented in Table~2. The
correlations are rather significant in all cases, being the best in
the case of charges $Q=0,1$ and antisymmetric field, in agreement with
\citet{Tinyakov:2001ir}. In this case, the data count
$N_d(2.7^{\circ})= 13$, while 2 is expected in average for a uniform
background.  The probability for this to occur by chance is
$p(\delta)=3\times 10^{-7}$ ($5.1\sigma$). The dependence of
$p(\delta)$ on $\delta$ in this case is shown in Fig.~\ref{SS-EV}.
% 
\if 0
%%%%%%%%%%%%%%%%%%%%%%%%%%%%%%%%%%%%%%%%%%%%%%%%%%%%%%%%%%%%%%%%%%%%
\begin{figure}
\plotone{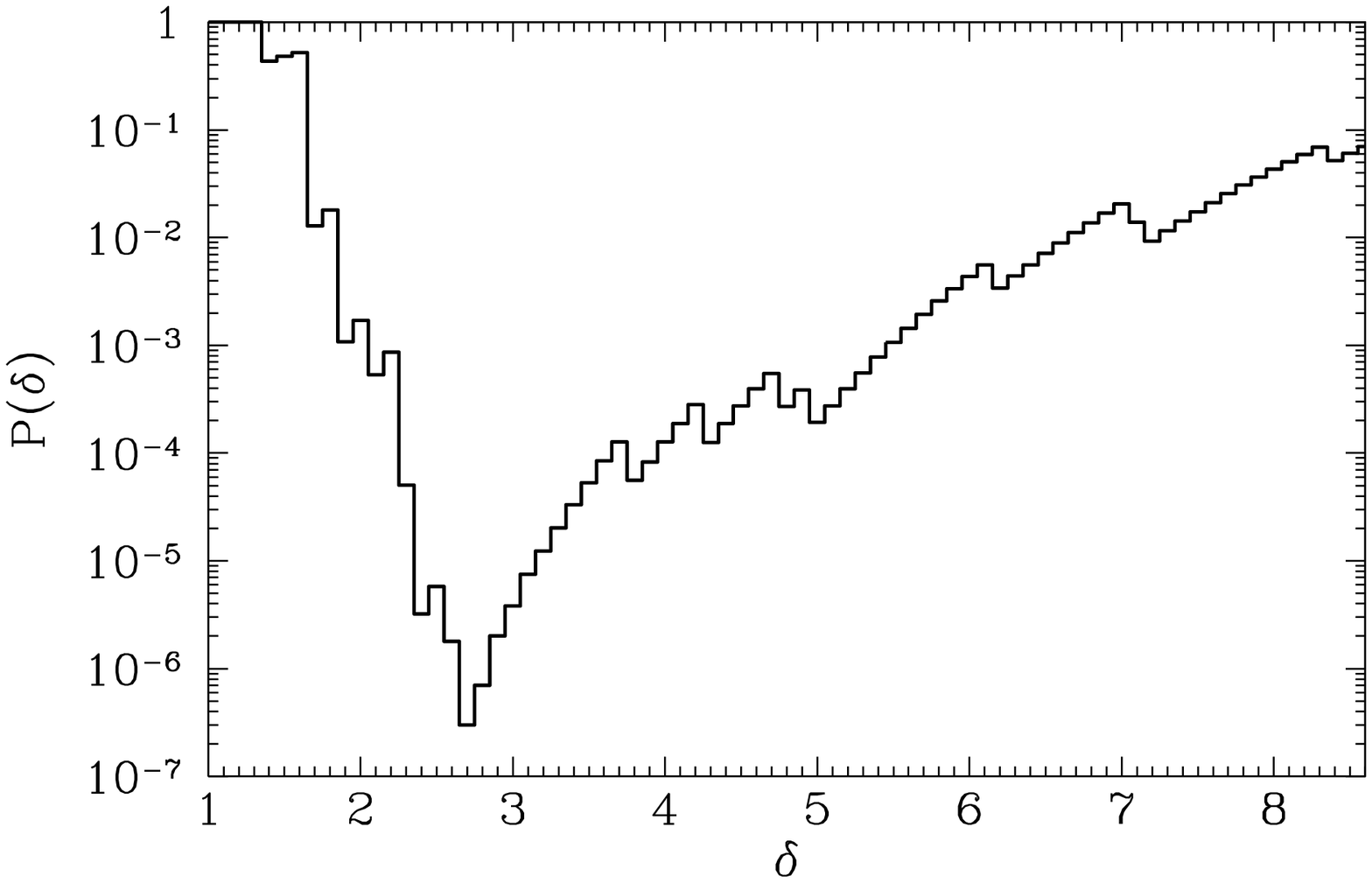}
\caption{Significance of correlations between 14 $\gamma$-ray loud BL
Lac objects and UHECR as a function of the angular scale $\delta$ for
the $Q=0,1$ charge composition. This corresponds to the lowest
probability entry of the Table 2.}
\label{SS-EV}
\end{figure}
%%%%%%%%%%%%%%%%%%%%%%%%%%%%%%%%%%%%%%%%%%%%%%%%%%%%%%%%%%%%%%%%%%%%
\fi 

{}From the analysis of Table~2 one is tempted to conclude that the
case $Q=0,1$ (neutral and positively charged particles) is favored.
However, present statistics are not enough for such a conclusion, as
the following simple argument shows. The data count is subject to
fluctuations that may be roughly estimated as $\pm\sqrt{N_d}$ (these
fluctuations would be observed if the AGASA and Yakutsk experiments
were repeated many times). If the ``average'' data count were 10,
counts from seven to 13 would occur equally often. Corresponding
probabilities $p(\delta)$ would range from $10^{-4}$ to
$10^{-7}$. Thus, unlike correlations themselves, the {\em difference}
between the cases $Q=0$, $Q=0,1$ and $Q=0,\pm 1$ can easily be
explained by a fluctuation.

Energies and charges of UHECR events that contribute into correlations
with $\gamma$-ray--loud BL Lac objects are listed in columns (7) and
(8) of Table~1 (the antisymmetric magnetic field model is
assumed). Multiple charges in column (8) mean that the corresponding
event contributes to correlations under different charge assignments.

The comparison between Table~1 of this Letter and Table~1 of
\citet{Tinyakov:2001nr} shows that the same BL Lac objects and
cosmic rays contribute to correlations in Ref.~\citep{Tinyakov:2001nr}
and in the case of $Q=0$ presented above. In \citet{Tinyakov:2001nr},
the set of brightest BL Lac objects was selected by imposing cuts on
redshift, apparent magnitude, and radio flux. In the resulting subset
of 22 BL Lac objects, five candidate sources were identified. It is
remarkable that four out of these five candidates, in particular all
three that correlate with UHECR multiplets, are among the 14 BL Lac
objects that comprise the intersection of BL Lac and EGRET catalogs,
$\gamma$-ray--loud BL Lac objects. Even more remarkable is that out
of 10 remaining BL Lac objects, four correlate with cosmic rays after
correction for GMF. Among the remaining six that do not correlate with
UHECRs, two objects are situated in the Southern hemisphere invisible
for Yakutsk and AGASA experiments. These objects can be excluded from
correlation analysis. Thus, the majority of $\gamma$-ray--loud BL Lac
objects (eight out of 12) correlate with UHECR. One concludes that the
ability to emit $\gamma$-rays may be used as the physical criterion that
allows to select actual UHECR sources from the set of all BL Lac
objects.

BL Lac objects are typically faint objects. Some of the unidentified
EGRET sources may be actually BL Lac objects that have not not yet
been observed at other wavelengths, or have been observed but not
identified as BL Lac objects. If this is the case and our conclusion
about the connection between $\gamma$-ray and UHECR emissions is
correct, one may expect correlations between unidentified EGRET
sources and UHECRs. To check this, we calculated correlations between
UHECRs and unidentified EGRET sources having Galactic latitude
$|b|>10^{\circ}$ (the cut $|b|>10^{\circ}$ is made to increase the
fraction of extragalactic sources as, according to \citet{Grenier},
the total number of such sources is expected to be 30-40 only). This
set contains 96 objects. Correlations are best when all particles are
assumed to have a charge of $Q=+1$; corresponding significance is
$p(\delta)= 10^{-4}$ at $\delta =3^{\circ}$. Table~3 summarizes EGRET
sources and cosmic rays that contribute to correlations. It is
interesting to note that, unlike Table~1, Table~3 seems to favor
positively charged particles. We expect that some EGRET sources listed
in Table~3 are BL Lac objects that have not yet been observed.

To summarize, there exists a significant correlation of arrival
directions of UHECRs with $\gamma$-ray--loud BL Lac objects (BL Lac
objects that may be associated with the EGRET sources). This confirms
the conjecture that strong $\gamma$-ray emission is a characteristic
feature of those BL Lac objects that are the sources of UHECR. Present
data are compatible with charges of primary particles $Q=0$, $Q=+1$,
$Q=0,+1$ and $Q=0,\pm 1$, although they favor the latter two cases. It
does not seem possible, with the present statistics, to distinguish
between these cases on the basis of correlation analysis, but it
should be possible in the future. This question is of particular
interest since specific charge composition is a good signature of most
of the existing models.  Charge $Q=0,\pm 1$ would speak strongly for
neutrino models. Charge $Q=1$ would favor protons accelerated in BL
Lac objects (note that energies of most of the $Q=1$ events in Table~1
would allow them to reach us from super-GZK distances provided
extragalactic magnetic fields are small). The cases $Q=0,1$ and $Q=0$
would suggest, in view of the distance to BL Lac objects and presence
of neutral particles, the existence of new physics (e.g., exotic
neutral particles \citep{Chung,Berezinsky2,Gorbunov} or violation of
Lorentz invariance \citep{Coleman,Dubovsky2}).

The results presented here suggest that the sources of UHECRs are
high-energy--peaked BL Lac objects located opposite the flat-spectrum
radio quasar end of the ``unified blazar
sequence''\citep{blazar-sequence}. This does not contradict the
conclusions of \citet{Sigl}, who found no correlations between UHECR
and {\em identified\/} EGRET blazars. Indeed, most of the latter are
high-polarization blazars, and not low-polarization,
high-energy--peaked BL Lac objects that, according to our study, are
the most probable sources of UHECR.

The work of P.T.\ is supported by the Swiss Science Foundation, grant
21-58947.99. The work of D.G.\ and S.T.\ is supported in part by the
program SCOPES of the Swiss National Science Foundation, project
No.~7SUPJ062239, by SSLSS grant 00-15-96626, and by RFBR grant
02-02-17398. The work of D.G., I.T., and S.T. is supported in part by
INTAS grants YSF 2001/2-142, 99-1065 and YSF 2001/2-129, respectively.
This Letter made use of the SIMBAD database, operated at the Centre de
Donn\'ees Astronomiques de Strasbourg, Strasbourg, France.

\newpage 

%%%%%%%%%%%%%%%%%%%%%%%%%%%%%%%%%%%%%%%%%%%%%%%%%%%%%%%%%%%%%%%%%%%%%%%%%
\begin{table}
\begin{tabular}{l|c|c|c|c|c|c|l}
\footnotesize 3EG J &\footnotesize E ID &\footnotesize Possible BLL      &\footnotesize l &\footnotesize b &\footnotesize z  &\footnotesize E &\footnotesize Q\\
\footnotesize (1) &\footnotesize (2) &\footnotesize (3) &\footnotesize (4) &\footnotesize (5) &\footnotesize (6) &\footnotesize (7) &  \footnotesize (8) \\
\hline
\footnotesize 0433+2908 &\footnotesize AGN &\footnotesize 2EG J0432+2910*   &\footnotesize  170.5 &\footnotesize  -12.6 &\footnotesize   ---  &\footnotesize  5.47 &\footnotesize 0,$\pm1$\\
\footnotesize  &\footnotesize &\footnotesize &\footnotesize  &\footnotesize   &\footnotesize     &\footnotesize   4.89  &\footnotesize 0,$+1$\\
\footnotesize 0808+5114 &\footnotesize AGN? &\footnotesize 1ES 0806+524*   &\footnotesize  166.2 &\footnotesize  32.91 &\footnotesize 0.138 &\footnotesize 3.4  &\footnotesize 0\\
\footnotesize  &\footnotesize  &\footnotesize    &\footnotesize      &\footnotesize  &\footnotesize  &\footnotesize 2.8  &\footnotesize 0\\
\footnotesize  &\footnotesize  &\footnotesize    &\footnotesize    &\footnotesize  &\footnotesize  &\footnotesize 2.5  &\footnotesize 0\\
\footnotesize 0812-0646 &\footnotesize AGN? &\footnotesize 1WGA J0816.0-0736 &\footnotesize  229.8 &\footnotesize  14.96 &\footnotesize  0.04 &\footnotesize --- &\footnotesize \\
\footnotesize 1009+4855 &\footnotesize AGN? &\footnotesize GB 1011+496     &\footnotesize  165.5 &\footnotesize  52.71 & \footnotesize   0.2 &\footnotesize --- &\footnotesize \\
\footnotesize 1052+5718 &\footnotesize AGN? &\footnotesize RGB J1058+564* &\footnotesize  149.6 &\footnotesize  54.42 &\footnotesize 0.144 &\footnotesize  7.76  &\footnotesize 0,$-1$\\
\footnotesize  &\footnotesize &\footnotesize &\footnotesize   &\footnotesize     &\footnotesize &\footnotesize  5.35  &\footnotesize 0,$-1$\\
\footnotesize  &\footnotesize &\footnotesize &\footnotesize   &\footnotesize     &\footnotesize &\footnotesize  5.50  &\footnotesize $-1$\\
\footnotesize 1222+2841 &\footnotesize AGN &\footnotesize ON 231*         &\footnotesize  201.7 &\footnotesize  83.29 &\footnotesize 0.102 &\footnotesize --- &\footnotesize \\
\footnotesize 1310-0517 &\footnotesize &\footnotesize 1WGA J1311.3-0521 &\footnotesize  312.1 &\footnotesize  57.16 &\footnotesize  0.16 &\footnotesize --- &\footnotesize \\
\footnotesize 1424+3734 &\footnotesize &\footnotesize TEX 1428+370   &\footnotesize  63.95 &\footnotesize  66.92 &\footnotesize 0.564 &\footnotesize  4.97  &\footnotesize 0,$+1$\\
\footnotesize 1605+1553 &\footnotesize AGN &\footnotesize PKS 1604+159*   &\footnotesize  29.38 &\footnotesize  43.41 &\footnotesize ---    &\footnotesize --- &\footnotesize \\
\footnotesize 1621+8203 &\footnotesize &\footnotesize 1ES 1544+820    &\footnotesize  116.5 &\footnotesize  32.97 &\footnotesize ---    &\footnotesize 2.7 &\footnotesize +1\\
\footnotesize 1733+6017 &\footnotesize &\footnotesize RGB J1742+597    &\footnotesize  88.46 &\footnotesize  31.78 &\footnotesize ---    &\footnotesize 2.5 &\footnotesize +1\\
\footnotesize  &\footnotesize &\footnotesize &\footnotesize      &\footnotesize  &\footnotesize     &\footnotesize 6.93 &\footnotesize $-1$\\
\footnotesize 1850+5903 &\footnotesize &\footnotesize RGB J1841+591    &\footnotesize  88.68 &\footnotesize  24.29 &\footnotesize  0.53 &\footnotesize 5.8 &\footnotesize +1\\
\footnotesize  &\footnotesize &\footnotesize  &\footnotesize     &\footnotesize  &\footnotesize  &\footnotesize 2.8 &\footnotesize +1\\
\footnotesize 1959+6342 &\footnotesize &\footnotesize 1ES 1959+650     &\footnotesize  98.0  &\footnotesize  17.67 &\footnotesize 0.047 &\footnotesize  5.5 &\footnotesize +1\\
\footnotesize 2352+3752 &\footnotesize AGN? &\footnotesize TEX 2348+360   &\footnotesize  109.5 &\footnotesize -24.91 &\footnotesize 0.317 &\footnotesize --- &\footnotesize \\
\end{tabular}
\caption{List of BL Lac objects associated with EGRET sources and
UHECR which contribute to correlations.}  {Note: (1) EGRET name; (2)
EGRET identification; (3) suggested BL Lac counterpart; the five
objects marked with as asterisk are the cases when suggested
identification argrees with the SIMBAD database; (4) and (5) Galactic
coordinates of the BL Lac counterpart; (6) redshift of the BL Lac
counterpart as given by \citet{Veron}; (7) energies of correlating
cosmic rays in units of $10^{19}$~eV; (8) UHECR charge assignments
under which the correlation occurs. }
\end{table}
%%%%%%%%%%%%%%%%%%%%%%%%%%%%%%%%%%%%%%%%%%%%%%%%%%%%%%%%%%%%%%%%%%%%%%%%%

%%%%%%%%%%%%%%%%%%%%%%%%%%%%%%%%%%%%%%%%%%%%%%%%%%%%%%%%%%%%%%%%%%%
\begin{table}
\begin{tabular}{l|c|c|c|c|c|c}
%\footnotesize 3EG J &\footnotesize E ID &\footnotesize Possible BLL
\footnotesize Q & 
\multicolumn{3}{c|}{\footnotesize antisymmetric field} & 
\multicolumn{3}{c}{\footnotesize symmetric field}  \\ 
\hline                 
       &\footnotesize  $p(\delta )$ &\footnotesize  $N_d(\delta)$  &\footnotesize  $\delta$ 
       &\footnotesize  $p(\delta )$ &\footnotesize  $N_d(\delta)$  &\footnotesize  $\delta$ \\
\hline
\footnotesize $0$    &\footnotesize  $ 10^{-4}$ &\footnotesize  8  &\footnotesize  $2.9^{\circ}$ 
       &\footnotesize  $ 10^{-4}$ &\footnotesize  8  &\footnotesize  $2.9^{\circ}$  \\
\footnotesize $+$    &\footnotesize  $7\cdot 10^{-5}$   &\footnotesize  8  &\footnotesize  $2.7^{\circ}$ 
       &\footnotesize  $9\cdot 10^{-4}$   &\footnotesize  9  &\footnotesize  $3.7^{\circ}$  \\
\footnotesize $0,+$  &\footnotesize  $3\cdot 10^{-7}$   &\footnotesize  13 &\footnotesize  $2.7^{\circ}$ 
       &\footnotesize  $2\cdot 10^{-6}$ &\footnotesize  12 &\footnotesize  $2.6^{\circ}$  \\
\footnotesize $0,\pm$&\footnotesize  $10^{-6}$ &\footnotesize  15 &\footnotesize  $2.8^{\circ}$ 
       &\footnotesize  $2\cdot 10^{-6}$ &\footnotesize  15 &\footnotesize  $2.9^{\circ}$
\end{tabular}
\caption{Summary of correlations between 14 BL Lac objects and 65
cosmic rays for different charge assignments and models of the GMF.}
\end{table}
%%%%%%%%%%%%%%%%%%%%%%%%%%%%%%%%%%%%%%%%%%%%%%%%%%%%%%%%%%%%%%%%%%%%%%%%%

%%%%%%%%%%%%%%%%%%%%%%%%%%%%%%%%%%%%%%%%%%%%%%%%%%%%%%%%%%%%%%%%%%%%%%%%%
\begin{table}
\begin{tabular}{l|c|c|c|l}
 \footnotesize 3EG J     &\footnotesize      l &\footnotesize      b &  \footnotesize E &  \footnotesize Q\\
% \footnotesize (1)       &\footnotesize (2) &\footnotesize      (3) &\footnotesize      (4) &  \footnotesize (5) \\
\hline
 \footnotesize 0245+1758 &\footnotesize 157.6  &\footnotesize -37.11 &\footnotesize 3.2   &\footnotesize +1\\ 
 \footnotesize 0329+2149 &\footnotesize 165.0  &\footnotesize -27.88 &\footnotesize 4.8   &\footnotesize +1\\ 
 \footnotesize 0429+0337 &\footnotesize 191.4  &\footnotesize -29.08 &\footnotesize 6.19   &\footnotesize 0,$+1$\\ 
 \footnotesize 1227+4302 &\footnotesize 138.6  &\footnotesize  73.33 &\footnotesize 4.3   &\footnotesize +1\\ 
 \footnotesize 1308+8744 &\footnotesize 122.7  &\footnotesize  29.38 &\footnotesize 3   &\footnotesize +1 \\
 \footnotesize 1337+5029 &\footnotesize 105.4  &\footnotesize  65.04 &\footnotesize 5.68   &\footnotesize +1\\ 
 \footnotesize 1621+8203 &\footnotesize 115.53 &\footnotesize  31.77 &\footnotesize 2.7&\footnotesize +1\\ 
 \footnotesize 1824+3441 &\footnotesize  62.49 &\footnotesize  20.14 &\footnotesize 9.79   &\footnotesize 0,$\pm1$ \\
 \footnotesize 1835+5918 &\footnotesize  88.74 &\footnotesize  25.07 &  5.8&\footnotesize +1\\ 
 \footnotesize  &\footnotesize   &\footnotesize   &  2.8&\footnotesize +1
\end{tabular}
\caption{List of unidentified EGRET sources correlating with cosmic
rays.}  {Note: Columns are the same as in Table 1.}
\end{table}
%%%%%%%%%%%%%%%%%%%%%%%%%%%%%%%%%%%%%%%%%%%%%%%%%%%%%%%%%%%%%%%%%%%%%%%

%%%%%%%%%%%%%%%%%%%%%%%%%%%%%%%%%%%%%%%%%%%%%%%%%%%%%%%%%%%
\begin{figure}
\plotone{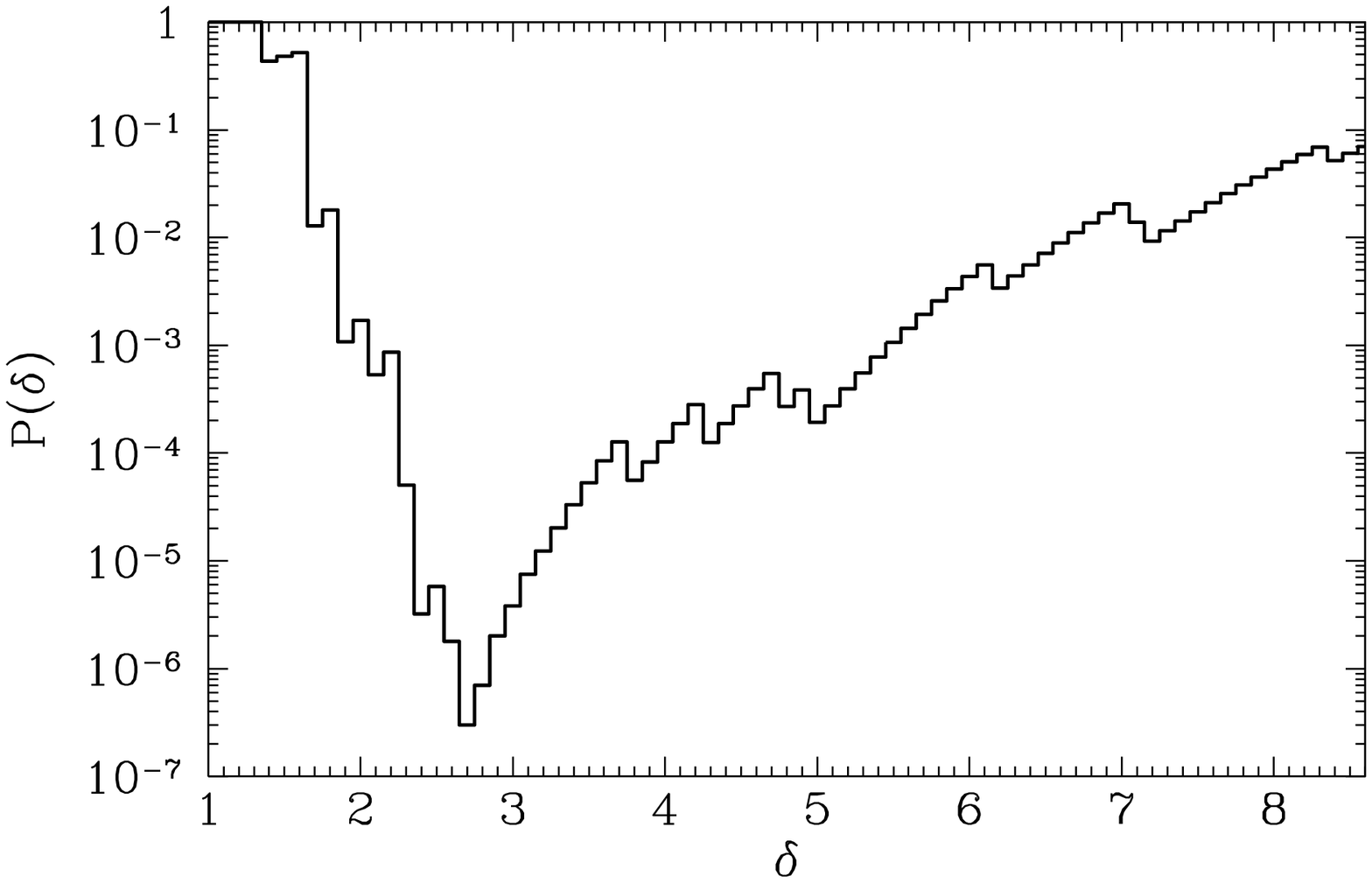}
\caption{Significance of correlations between 14 $\gamma$-ray--loud BL
Lac objects and UHECRs as a function of the angular scale $\delta$ for
the $Q=0,1$ charge composition. This corresponds to the lowest
probability entry of Table 2.}
\label{SS-EV}
\end{figure}
%%%%%%%%%%%%%%%%%%%%%%%%%%%%%%%%%%%%%%%%%%%%%%%%%%%%%%%%%%%%

\end{document}